\documentclass{article}

\usepackage[english]{babel}

\usepackage[letterpaper,top=2cm,bottom=2cm,left=3cm,right=3cm,marginparwidth=1.75cm]{geometry}

\usepackage{amsmath}
\usepackage{graphicx}
\usepackage[colorlinks=true, allcolors=blue]{hyperref}
\usepackage{mhchem}
\usepackage{soul,color}
\usepackage{dcolumn}
\usepackage{bm}

\title{A Novel Buckle-Free Large Rib Microdisk with Sub-Micron Thickness}
\author{Shahin Honari and Saeed Farajollahi and Tao Lu}

\begin{document}
\maketitle

\begin{abstract}
Thin large microdisks, that are key for dense spectral microcomb generation at visible to UV wavelengths, face challenges in fabrication. One of the most difficult issues is the buckling effect that significantly reduces the cavity optical quality factor. This work introduces a novel rib disk structure that significantly mitigates the buckling effects. Using this approach, we obtained millimeter size buckle-free microdisks with sub-micron thickness and high optical quality factor exceeding $10^7$.
\end{abstract}

\section{Introduction}

Microcavities with high opitcal quality factors (Q) have been one of the most sought after optical structures for many years. With applications ranging from biosensing and nanoparticle detection to quantum electrodynamics and comb generation, these devices have been at the forefront of many scientific endeavours~\cite{vahala2003optical,Arnold08,baaske2014single,PhysRevLett.108.120801,lu2011high,Yu_Lu_Cavity_Molecules,kippenberg2011microresonator,papp2014microresonator,kippenberg2008cavity}. Among the materials normally used to make microcavities, \ce{SiO_2} offers unique advantages. The low loss nature of the thermally grown oxide on top of  silicon, and compatibility of the platform with already existing CMOS fabrication technologies, made SoS (Silica-on-Silicon) platform a great candidate for optical devices fabrication~\cite{jalali2006silicon}.

One of the fabrication challenges that reduces the applications of SoS platforms, is the stress build up at the interface between silica and silicon~\cite{doi:10.1063/1.4789370}. The stress arises from the fact that the two materials have different thermal expansion coefficients. Hence, when cooling down from high temperatures (exceeding 1,000~$^\circ$C when growing the oxide) to room temperature, significant stress will built up at the oxide interface~\cite{10.1115/1.3119508}. This hinders microdisk fabrication by buckling the edges of the disk at high undercuts and deteriorates the quality factor of the disk. This issue is more severe in large disks, which are necessary for dense microcomb generation. Although large ultra-high Q microdisks have been demonstrated without significant buckling problem, these devices were fabricated using thick oxide layers, accompanied with tedious high temperature annealing cycles to remove the stress~\cite{lee2012chemically,wu2020greater,doi:10.1063/5.0051674}. However, there is an increasing interest for thin oxide microdisks, specifically for comb generation applications at shorter wavelengths. Short wavelength comb generation is achievable with thinner microdisks~\cite{lee2017towards}. Multi-layer oxide disks with different wedge angle have been proposed before to control the dispersion~\cite{yang2016broadband}, but a fabrication procedure for a large thin disk is still missing. It has been shown that, for soliton microcomb generation, the needed anomalous dispersion at shorter wavelengths can only be satisfied with thin oxides~\cite{lee2017towards}. Thus, the need for thin oxide microdisks cannot be overstated. Our goal in this work is to fabricate buckle-free ultra-high Q large disk with thin oxide, by introducing a new rib geometry. The approach introduced in this work can address the buckling challenge without limiting the size and the thickness of the disk, and does not require any complicated post processing. Consequently, the fabricated device maintains high optical quality factor. This, to the best of our knowledge, is the first time such remedy is demonstrated to solve the buckling problem in large thin microdisks.

\section{Discussion}
\begin{figure*}[h]
\centering
\includegraphics[width=1.0\linewidth]{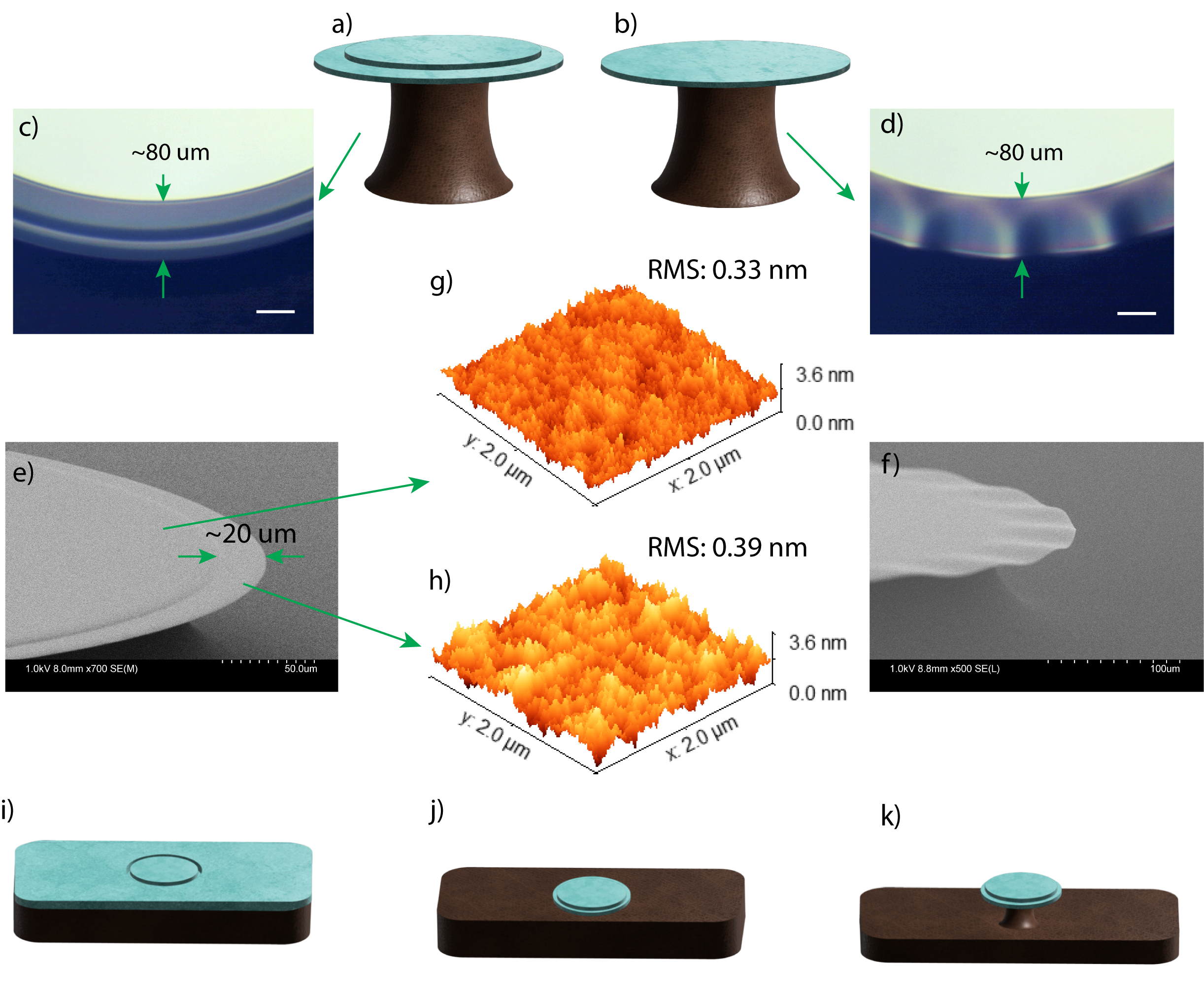}
\caption{A 3-D model of a) the proposed rib disk and b) a regular disk. The top view micrograph c) shows no buckling on the rib disk edge while on d) the regular disk with the same undercut ratio is severely buckled. Plot e) further shows the top view SEM of the rib disk. The width of the disk rib is around~$20~\mu$m. f) is the SEM image of a buckled regular disk. The AFM images at g) and h) show the surface roughness of the top part and the rib part of the rib disk respectively, RMS values obtained for each part is shown in the figure. Roughness RMS is 0.33~nm for the top part and 0.39~nm for the rib part.  i), j) and k) are the fabrication steps of the rib disk. The white bars on optical microscope images are $50~\mu$m each.}
\label{Figure_SEM_AFM}
\end{figure*}
\begin{figure}[htbp]
\centering
\includegraphics[width=0.8\textwidth]{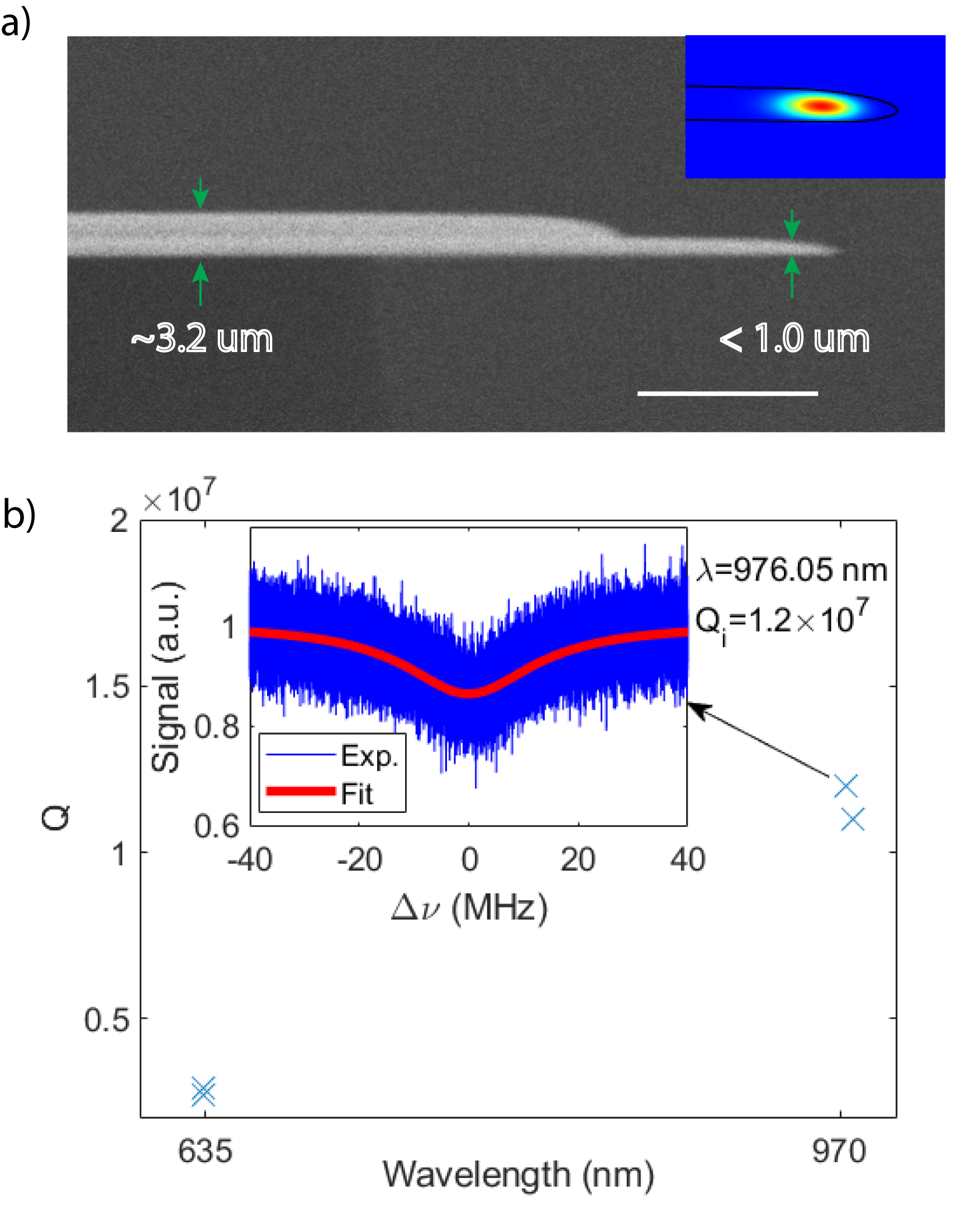}
\caption{a) A sideview SEM image of a rib disk along with the simulated mode profile (inset), The thickness of the rib and top part of the disk were calculated after accounting for a slight slope of the sample. The graph at b) represents optical Q data for two different wavelengths with the normalized transmission spectrum in the inset. The white bar in the SEM image is $20~\mu$m in length.}
\label{Figure_Q}
\end{figure}

The proposed rib disk structure is shown in Fig.~\ref{Figure_SEM_AFM}a, the thin silica layer at the edge of the disk is supported by a thicker top part with slightly smaller diameter, which provides sufficient mechanically strength to prevent thin edge from buckling. In our experiment, we fabricated a $1$~mm-diameter disk on a SoS wafer with a $4~\mu$m thick silica thin film. After a chemo-mechanical polishing (see fabrication details in later section), the thickness of the top part is about $3.2~\mu$m while the rib edge thickness reduces to below $1.0~\mu$m (see Fig.~\ref{Figure_Q}a). The radius difference between the outer ring and the top support (the rib width) is $20~\mu$m to $30~\mu$m (Fig.~\ref{Figure_SEM_AFM}e). As shown by the microscope picture in Fig.~\ref{Figure_SEM_AFM}c and SEM micrograph Fig.~\ref{Figure_SEM_AFM}e, no evidence of buckling was observed even we undercut the silicon pillar as far as $80~\mu$m. In contrast, a $1~$mm-diameter regular disk fabricated on a $2~\mu$m silica thin film\footnote{The disk thickness is around $1.5~\mu$m after polishing.} (Fig.~\ref{Figure_SEM_AFM}b) with the same undercut displays severe buckling as shown in the microscope image (Fig.~\ref{Figure_SEM_AFM}d) and SEM image(Fig.~\ref{Figure_SEM_AFM}f). Further, we measured the surface roughness of the rib disk using an atomic force microscope (AFM) both at the top and at the rib to see the smoothness of the surface. As shown in Fig.~\ref{Figure_SEM_AFM}g-h, the surface roughness RMS value of the top is slightly lower, and is about 0.33~nm , whereas the rib part has RMS value of 0.39~nm for surface roughness. This small discrepancy can be attributed to the disk geometry before polishing, where the top surface has more contact with the polishing pad due to the height difference between the top and the rib. It is worth mentioning that both roughness RMS values are close to what was obtained in previous works~\cite{doi:10.1063/5.0051674}. The AFM data was obtained on a $2~\mu$m${\times}2~\mu$m surface since the rib is not wide enough for large area scans. Such smooth surface in combination to the absence of the buckles are key to maintain high Q of such large thin disk.

The fabrication process starts with defining a ring on a $4~\mu$m grown silica on silicon wafer. In order to do so, a Raith 50 e-beam writer is used to define a ring pattern with desired width and diameter, in this case $980~\mu$m inner diameter and $40~\mu$m width, on PMMA resist, followed by a developing step using MIBK solution. Then the wafer is wet etched using a buffered HF solution (Transene), to etch away the ring to about $1~\mu$m. The PMMA resist is then washed away with acetone~(Fig.~\ref{Figure_SEM_AFM}i). The next step is to define the disk on top of the ring. Note that the disk has to partially cover the ring. The overlap region between the two patterns defines the size of the rib. Disks are made using a photolithography step followed by a HF wet etch all the way down to the silicon substrate, and a polishing process to reduce the surface roughness of the disk and boost the Q~(Fig.~\ref{Figure_SEM_AFM}j). The details of the polishing process has been discussed in~\cite{doi:10.1063/5.0051674}. The final step is using \ce{XeF_2} to etch the silicon underneath the disk and form the pillars~(Fig.~\ref{Figure_SEM_AFM}k). Buckling could happen in this last step by reducing the \ce{Si/SiO_2} interface, and releasing the stress. SEM sideview images and profilometer data suggests the thickness of the disk at the edges is about $980$~nm to $1~\mu$m.

We then measured the Q at two different wavelengths using a reference interferometer setup described in Ref.~\cite{doi:10.1063/5.0051674}. The disk transmission spectrum is shown as the blue trace of Fig.~\ref{Figure_Q}b inset, which is fitted to a Lorentzian function (red trace of Fig.~\ref{Figure_Q}b inset) to obtain the intrinsic Q of the cavity. As shown in Fig.~\ref{Figure_Q}b (blue cross markers), the Q as high as $1.2{\times}10^7$ was achieved at 970~$\mu$m and ${2{\times}10^6}$ when measuring with a $635~\mu$m laser for this rib disk. The Q degradation at shorter wavelengths can be attributed to more surface scattering at these wavelengths~\cite{borselli2004rayleigh}, which may be improved through a refined chemo-mechanical polishing procedure. In contrast, we couldn't observe any resonance of the buckled regular disk as these buckles significantly reduce the photon lifetime inside the cavity.

\begin{figure}[ht]
\centering
\includegraphics[width=1.0\textwidth]{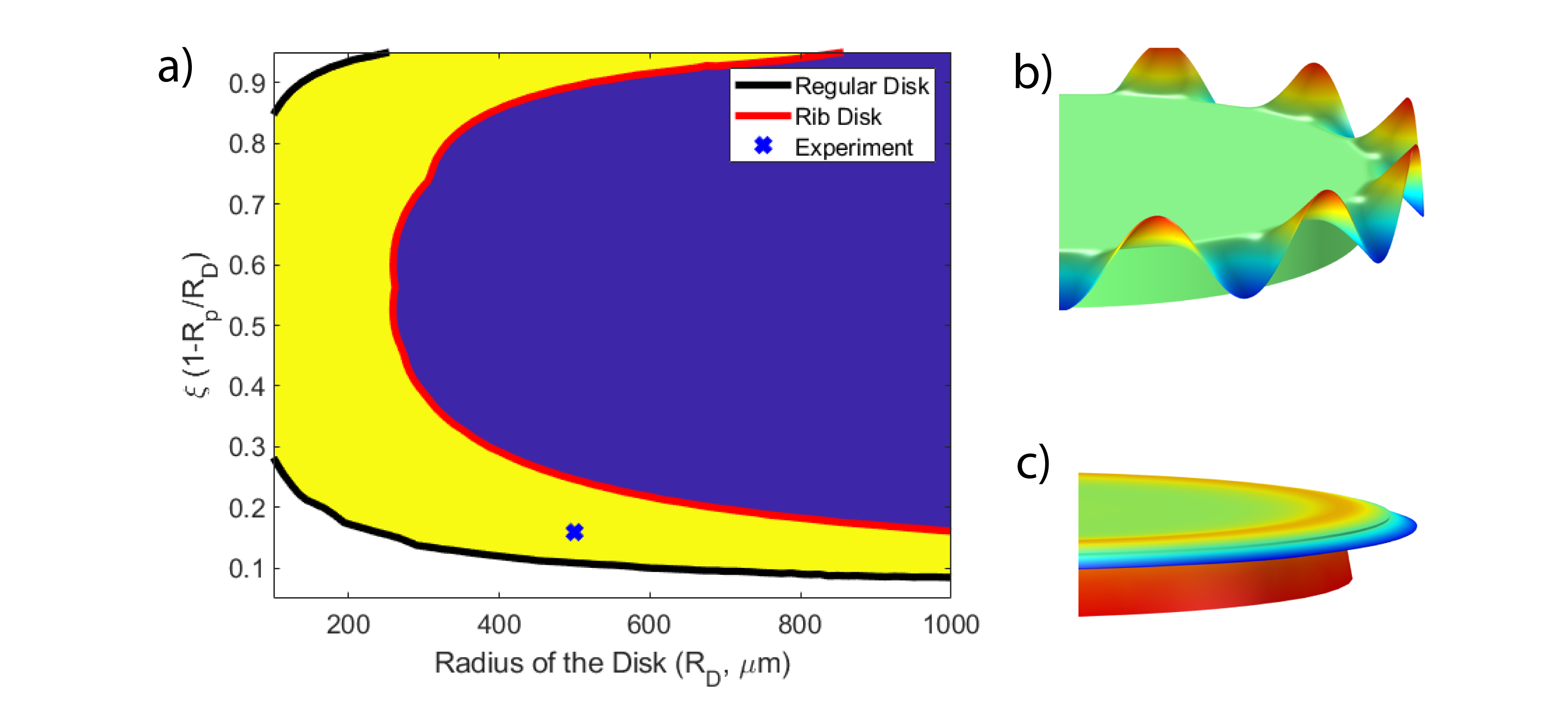}
\caption {a) Buckling threshold for rib and regular disk as a function of disk radius ($R_D$) and the undercut ratio $\xi$. Yellow coloured region in the graph represents the gain in our design, where the rib disk remains mechanically stable, but the regular disk with the same dimensions and undercut buckles, while in the blue coloured region both structures show buckling. At the setting ("x" marker), where both disks were fabricated, the simulation b) shows that a regular disk will buckle while a rib disk showing in c) will not buckle.}
\label{Figure_simulation}
\end{figure}
To further investigate the mitigation effectiveness, we did mechanical buckling simulations to our rib disk and a regular disk with the same thickness and diameter. To model the mechanical stability of our design, the thermal stress induced by different thermal expansion coefficients in silica and in silicon is simulated in COMSOL Multiphysics via finite element method. The linear buckling study in COMSOL is used to derive the critical load factor (CLF). CLF is defined as the ratio of the minimum load for buckling to the existing load. A CLF less than 1 means that the structure would face instability under thermal stress. The buckling threshold derived analytically in~\cite{doi:10.1063/1.4789370} now can be calculated numerically by doing a 2D parameter sweep on disk radius $R_D$ and the undercut ratio $\xi=1-R_p/R_D$ defined as 1 minus the ratio between the pillar $(R_p)$ and disk $(R_D)$ radii. It is shown that a boundary between buckled and unbuckled region can be obtained using numerical simulation on linear buckling for any thickness. This boundary, in our simulations, is defined by the contour of CLF=1.

Our simulation in Fig.~\ref{Figure_simulation}a displays the unity CLF contour of our rib disk (red curve) and a regular disk with the same disk thickness of $1~\mu$m (black curve). In general, the left-hand side of the curve shows critical load factors higher than one, which means the disk remains unbuckled. The right-hand side of the curve shows critical load factors below one, hence the disks will buckle.  Consequently, both rib and regular disks in the blue region will buckle. The yellow region in~Fig.~\ref{Figure_simulation}(a) shows the area in which the rib disk remains stable, but a regular disk will buckle. Further, in the white region neither the regular disk nor the rib disk will buckle.  The cross marker represents the configuration ($R_D=0.5$~mm and $\xi=0.17$) of the regular and rib disk we fabricated (also see Fig.~\ref{Figure_SEM_AFM}c and ~\ref{Figure_SEM_AFM}d). As shown, although the normal disk with the size and thickness specified will buckle (Fig.~\ref{Figure_simulation}b), our rib disk with same size will remain mechanically stable even at high undercuts (Fig.~\ref{Figure_simulation}c). These simulations are perfectly in par with the experimental results shown earlier.

\section{Conclusions}

In conclusion, we demonstrated novel buckle-free rib microdisks with high quality factors. Using the developed fabrication procedure, a 1-mm-diameter and $<1~\mu$m thick rib disk at an Q above ten million was demonstrated. With a refined fabrication, our simulation predicts that larger and thinner buckling free high Q disk can be made.  In future research, such rib disk will be developed to dense soliton microcombs at visible wavelengths that may enable in-vitro label free single molecule spectroscopy. In addition, the rib structure can be directly adopted to other material platforms such as \ce{SiN}, \ce{LiNbO3}, etc. to make large disk with enhanced mechanical rigidity. More importantly, as a novel cavity structure, many photonic properties are yet to be explored.

\section{Backmatter}

 This work was supported in part by the Nature Science and Engineering Research Council of Canada (NSERC) Discovery (Grant No. RGPIN-2020-05938),  and Threat Reduction Agency (DTRA) Thrust Area 7, Topic G18 (Grant No.GRANT12500317). We would like to acknowledge CMC Microsystems for the provision of products and services that facilitated this research, including the use of COMSOL for the numerical analysis.

 The authors acknowledge and appreciate the services provided by AMF and CAMTEC facilities at the University of Victoria. We specially thank Dr Elaine Humphrey and Mr. Jon Rudge for their help and valuable discussions.


\end{document}